\documentclass[11pt]{article}
\PassOptionsToPackage{authoryear}{natbib}
\usepackage[preprint,nonatbib]{neurips_2024}
\usepackage{hyperref} 
\usepackage{authblk} 
\usepackage{booktabs}
\usepackage{graphicx}
\usepackage[caption=false,font=footnotesize]{subfig}
\usepackage{amsfonts}
\usepackage{multicol}
\usepackage{tabularx}
\usepackage{enumitem,amsmath,bm,amssymb}
\usepackage{pifont,makecell}
\usepackage{tikz}
\usepackage{CJKutf8}

\usepackage{multirow}
\usepackage{cite}
\usepackage{comment}

\usepackage{xcolor}
\usepackage{enumitem}

\title{CosyVoice 3: Towards In-the-wild Speech Generation via Scaling-up and Post-training}

\author{
  \bf Zhihao Du, Changfeng Gao, Yuxuan Wang, Fan Yu, Tianyu Zhao, Hao Wang, Xiang Lv, Hui \\
  \bf Wang, Chongjia Ni, Xian Shi, Keyu An, Guanrou Yang, Yabin Li, Yanni Chen, Zhifu Gao,\\
  \bf Qian Chen, Yue Gu, Mengzhe Chen, Yafeng Chen, Shiliang Zhang, Wen Wang, Jieping Ye
}

\affil{Speech Team, Tongyi Lab, Alibaba Group}
\affil{\texttt{\{neo.dzh, funaudiollm\}@alibaba-inc.com}}
\date{}
\begin{document}

\maketitle

\begin{abstract}

In our prior works, we introduced a scalable streaming speech synthesis model, CosyVoice 2, which integrates a large language model (LLM) and a chunk-aware flow matching (FM) model, and achieves low-latency bi-streaming speech synthesis and human-parity quality. Despite these advancements, CosyVoice 2 exhibits limitations in language coverage, domain diversity, data volume, text formats, and post-training techniques. In this paper, we present \textbf{CosyVoice 3}, an improved model designed for \textbf{zero-shot multilingual speech synthesis in the wild}, surpassing its predecessor in content consistency, speaker similarity, and prosody naturalness. Key features of CosyVoice 3 include: 1) A \textbf{novel speech tokenizer} to improve prosody naturalness, developed via supervised multi-task training, including automatic speech recognition, speech emotion recognition, language identification, audio event detection, and speaker analysis. 2) A \textbf{new differentiable reward model for post-training} applicable not only to CosyVoice 3 but also to other LLM-based speech synthesis models. 3) \textbf{Dataset Size Scaling}: Training data is expanded from ten thousand hours to one million hours, encompassing 9 languages and 18 Chinese dialects across various domains and text formats. 4) \textbf{Model Size Scaling}: Model parameters are increased from 0.5 billion to 1.5 billion, resulting in enhanced performance on our multilingual benchmark due to the larger model capacity. These advancements contribute significantly to the progress of speech synthesis in the wild. We encourage readers to listen to the demo at \url{https://funaudiollm.github.io/cosyvoice3}.

\end{abstract}

\section{Introduction}

With the rapid development of generative neural networks, text-to-speech (TTS) synthesis has made significant progress, surpassing traditional concatenative and parametric methods in terms of synthetic quality \cite{DBLP:conf/interspeech/WangSSWWJYXCBLA17, DBLP:conf/icassp/ShenPWSJYCZWRSA18, DBLP:journals/corr/abs-1710-07654, DBLP:conf/iclr/PingPC19, DBLP:conf/nips/RenRTQZZL19, DBLP:conf/aaai/Li0LZL19, DBLP:conf/iclr/0006H0QZZL21}. In particular, zero-shot TTS models, which leverage vast multi-speaker datasets, can clone the timbre, prosody, and style of any speaker, and demonstrate performance superior to specific speaker TTS models, achieving human-like prosody naturalness and audio quality~\cite{DBLP:journals/corr/abs-2301-02111}.

Currently, zero-shot TTS models can be broadly categorized into three types: those using large language models (LLMs) to model discrete acoustic tokens \cite{DBLP:journals/corr/abs-2301-02111,DBLP:journals/tacl/KharitonovVBMGP23,DBLP:journals/corr/abs-2401-07333,DBLP:journals/corr/abs-2401-14321,DBLP:journals/corr/abs-2404-03204,DBLP:journals/corr/abs-2406-05370,DBLP:journals/corr/abs-2406-07855, DBLP:journals/corr/abs-2409-00750, DBLP:journals/corr/abs-2407-08551,wang2025spark}, those based on diffusion models to automatically learn internal alignments between speech and text \cite{DBLP:conf/nips/LeVSKSMWMAMH23,DBLP:conf/icml/JuWS0XYLLST000024,DBLP:conf/icassp/GuoDM0024,DBLP:conf/icassp/MehtaTBSH24, DBLP:conf/asru/GaoMZC23,DBLP:journals/corr/abs-2406-11427, DBLP:journals/corr/abs-2406-18009,DBLP:journals/corr/abs-2410-06885,DBLP:journals/corr/abs-2406-02430}, and coarse-to-fine hybrid systems that use auto-regressive LLMs to model coarse semantics, followed by non-autoregressive models such as diffusion models to render detailed speech features~\cite{DBLP:journals/corr/abs-2406-02430,cosyvoice,song2024touchtts,DBLP:journals/corr/abs-2409-03283,du2024cosyvoice2,jia2025ditar,deng2025indextts}. Considering the trade-offs between synthesis quality, streaming compatibility, and flexibility, such two-stage hybrid systems have become a mainstream choice in industrial applications. In our previous work, we developed CosyVoice 2~\cite{du2024cosyvoice2}. Through optimizing semantic token utilization, initializing with text-based LLMs, designing a bidirectional streaming scheme, and unifying instruction capability modeling, CosyVoice 2 achieves synthesis quality comparable to human speech, along with ultra-low latency bidirectional streaming synthesis capability that is virtually lossless~\cite{du2024cosyvoice2}.

\begin{figure}[t!]
    \centering
    \subfloat[Content Consistency]{\includegraphics[width=0.8\linewidth]{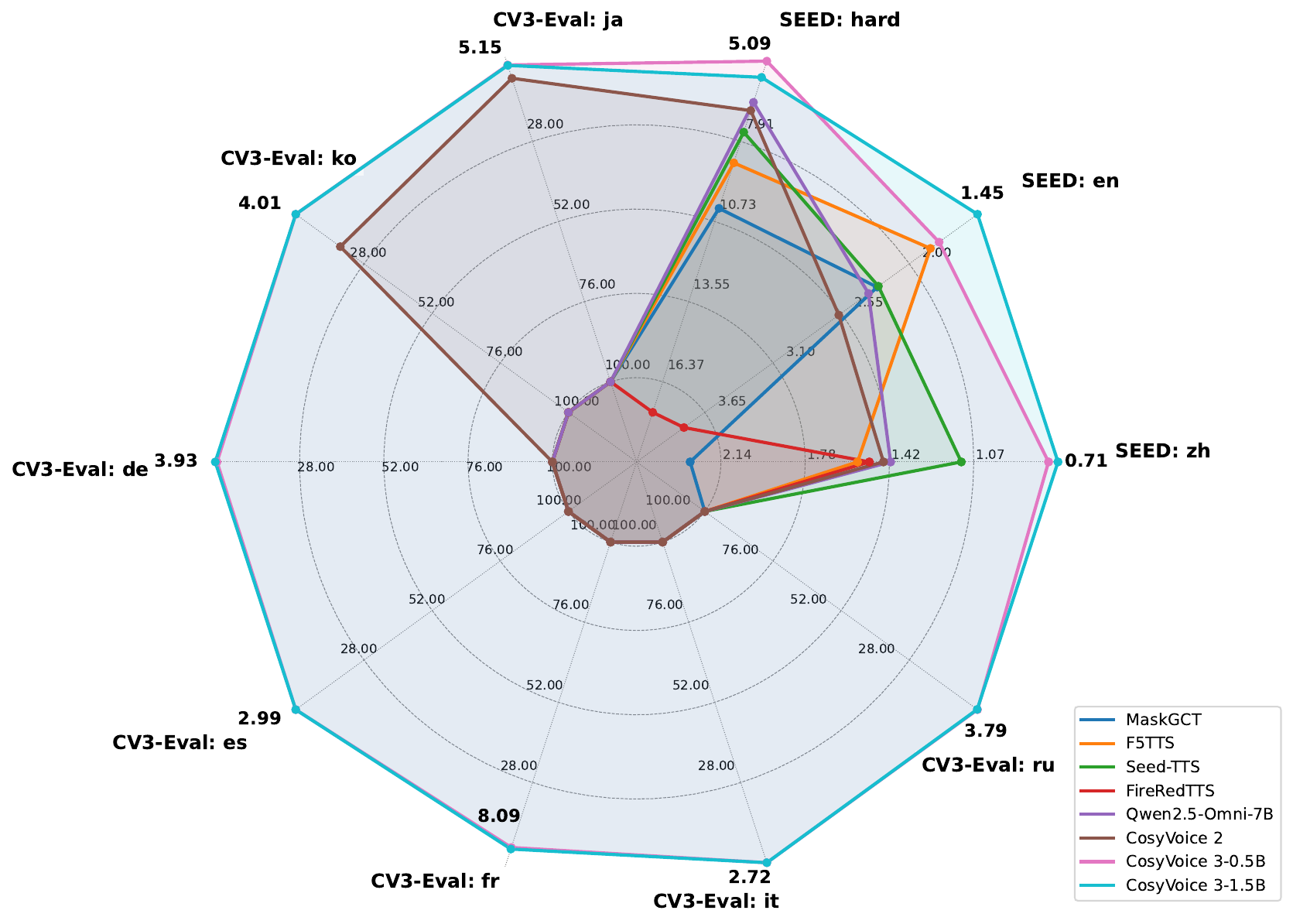}
    \label{fig:cc_radar}}\\
    \subfloat[Speaker Similarity]{\includegraphics[width=0.8\linewidth]{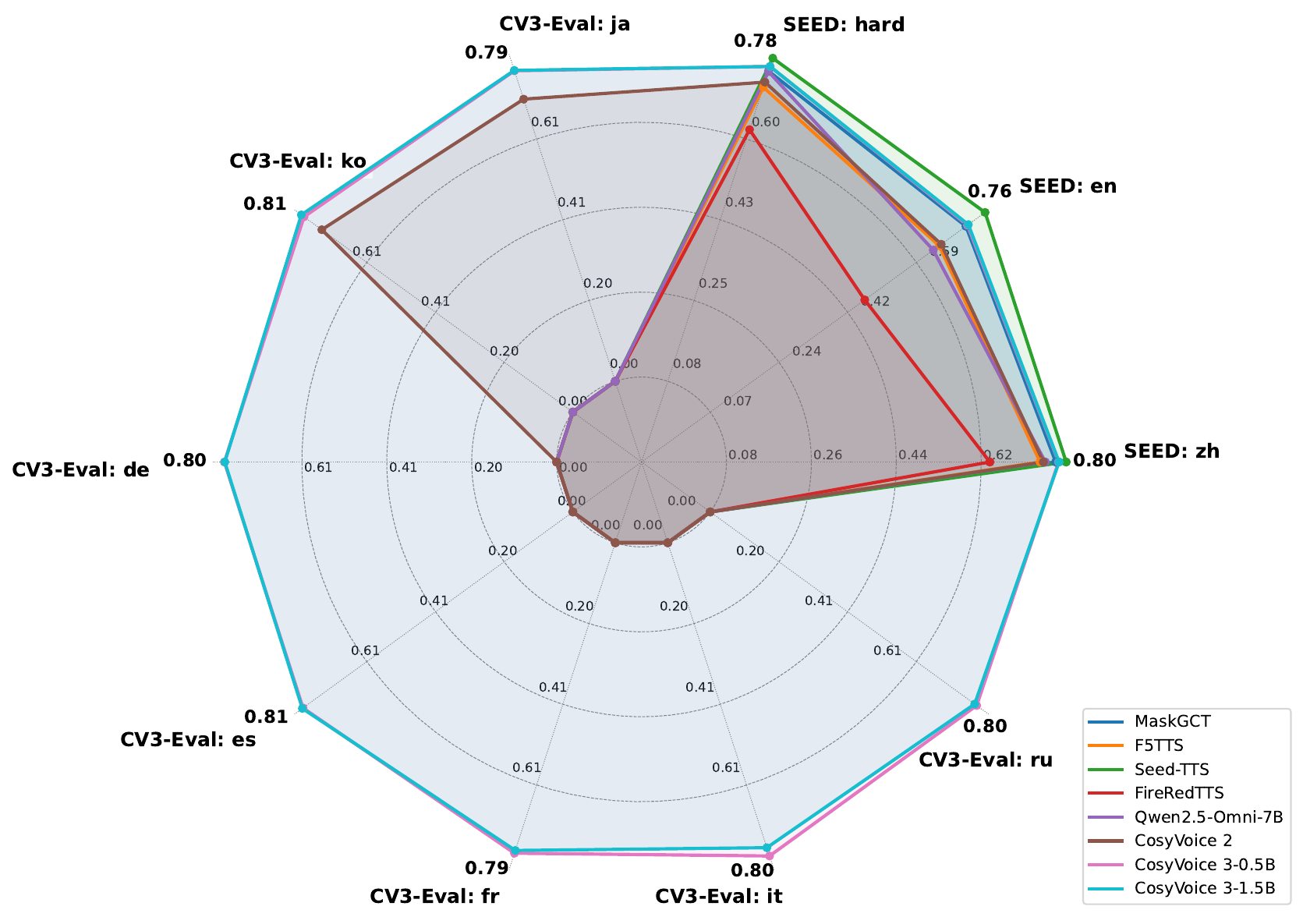}
    \label{fig:ss_radar}}
    \caption{Performance comparison between our CosyVoice 3 and competitive speech generation models in terms of content consistency and speaker similarity on various benchmarks. The numbers in (a) content consistency are CERs or WERs measured by ASR models. The numbers in (b) speaker similarity are cosine similarities of WavLM embeddings between reference and generated utterances. The error rates of 100.00 and the similarities of 0.00 mean that the released models do not support the languages.}
    \label{fig:benchmark}
\end{figure}

Although CosyVoice 2 performs well in general Chinese and English broadcast scenarios, it has noticeable limitations in language coverage, domain diversity, data volume, and text format variety, leaving significant room for improvement towards achieving in-the-wild speech generation. Furthermore, the scaling laws for models and data, as well as post-training techniques suitable for speech generation models, have not been thoroughly explored. To address these issues, we introduce CosyVoice 3, a large zero-shot speech generation model designed for in-the-wild applications, covering more languages and diverse scenarios, and significantly surpassing its predecessor CosyVoice 2 in content consistency, speaker similarity, and prosody naturalness. Our contributions can be summarized as follows:
\begin{itemize}[leftmargin=*]
\item We propose a \textbf{novel speech tokenizer} derived from a large audio understanding language model. Through supervised multi-task training, such tokenizer enables discrete speech tokens to better capture paralinguistic information such as emotion and pronunciation style.
\item We explore post-training strategies suitable for speech generation models and propose a new \textbf{differentiable reward optimization (DiffRO)} method, applicable not only to the CosyVoice series but also to other discrete-token-based speech synthesis models.
\item We validate \textbf{dataset size scaling} in the speech generation domain, expanding the training data from ten thousand hours to one million hours, covering 9 common languages, 18 Chinese accents/dialects, and various text formats, supporting better cross-lingual voice cloning. We also demonstrate the impact of \textbf{model size scaling} by increasing the model size from 0.5B to 1.5B, further enhancing the prosody naturalness.
\item To address the challenges of diversity and generalizability from unrestrained real-world speech synthesis scenarios, we release the \textbf{CV3-Eval} benchmark for zero-shot speech synthesis in the wild, which is built on authentic in-the-wild reference speech from Common Voice, FLUERS, EmoBox, and Web-crawled real-world audio data, and spans a broad range of languages and dialects, domains and environments, emotions and styles.
\end{itemize}
Through these improvements, CosyVoice 3 achieves state-of-the-art (SOTA) results on multiple benchmarks. We believe that CosyVoice 3 represents a solid step towards in-the-wild speech synthesis.

\section{CosyVoice 3}
\begin{figure}[thb]
    \centering
    \includegraphics[width=1.0\textwidth]{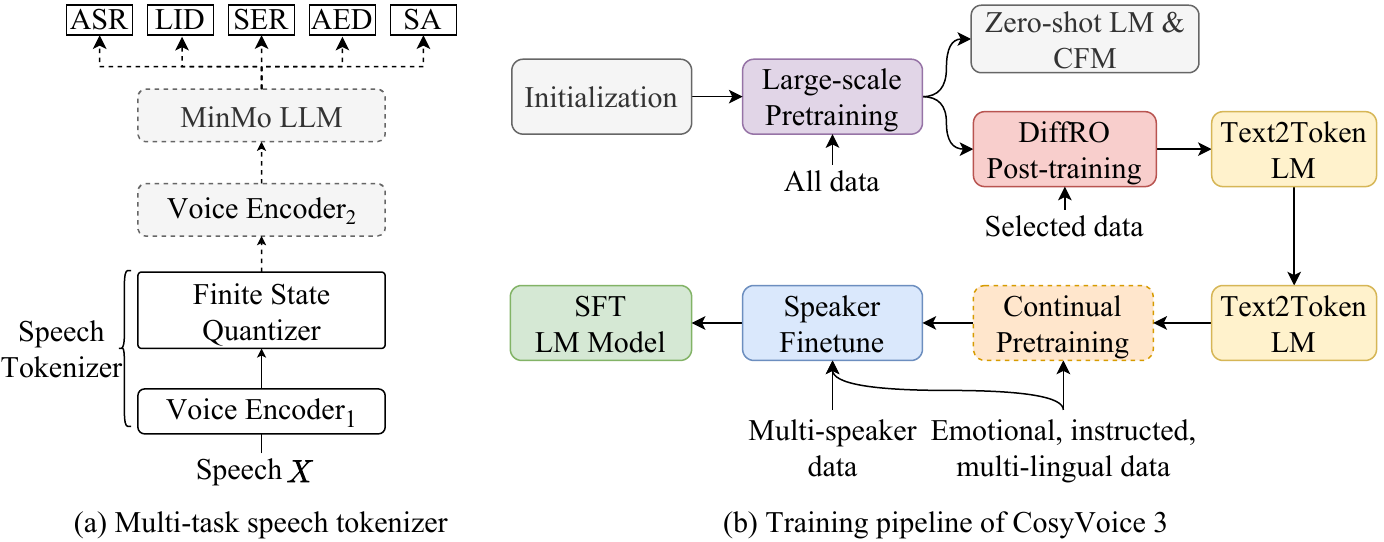}
    \caption{Illustrations of (a) Supervised multi-task trained speech tokenizer and (b) The training pipeline in CosyVoice 3. Modules with dashed boxes are only used in the training stage. The speech tokenizer is supervised trained on ASR, language identification (LID), speech emotion recognition (SER), audio event detection (AED), and speaker analysis (SA) tasks. CFM denotes the conditional flow matching model.}
    \label{fig:overview}
\end{figure}

Figure \ref{fig:overview} illustrates the training procedures for both the supervised multi-task supervised speech tokenizer and the generation models of CosyVoice 3. Different from its predecessor CosyVoice 2, the speech tokenizer in CosyVoice 3 is based on MinMo\cite{chen2025minmo}, a pretrained large-scale speech understanding model demonstrating strong performance across various speech tasks~\cite{chen2025minmo}. We also provide an overview of the training pipeline for the zero-shot and speaker fine-tuned (SFT) models, encompassing large-scale pretraining, post-training, continual pretraining, and multi-speaker fine-tuning. The post-training phase is aimed at surpassing the performance limitations of the training data, while the continual pretraining phase focuses on transferring capabilities, such as instruction controllability and multilingual synthesis, from the zero-shot model to the SFT models.

\subsection{Speech Tokenizer via Supervised Multi-task Training}
As shown in Figure \ref{fig:overview}a, different from CosyVoice 2 that inserts the Finite Scalar Quantization (FSQ) module~\cite{DBLP:conf/iclr/MentzerMAT24} into the encoder of the SenseVoice-Large ASR model \cite{funaduiollm}, for CosyVoice 3,  we insert the FSQ module into the voice encoder of the MinMo model \cite{chen2025minmo}. Compared to SenseVoice-Large ASR model, MinMo is an advanced multimodal LLM trained on an extensive dataset of over 1.4 million hours of speech, and showcases superior and SOTA performance across diverse benchmarks, including spoken dialogue, multilingual speech recognition, and emotion recognition.
To further enhance the ability of capturing semantic information, we leverage a subset of the training data for MinMo to conduct supervised multi-task learning for our speech tokenizer about 530,000 hour, including tasks such as multilingual ASR, language identification (LID), speech emotion recognition (SER), audio event detection (AED), and speaker analysis (SA).

During the training stage, the input speech $X$ goes through the $\mathrm{Voice\ Encoder}_1$ in Figure \ref{fig:overview}a to obtain the intermediate representations $H$, where $\mathrm{Voice\ Encoder}_1$ consists of 12 Transformer blocks with rotary positional embedding (RoPE)~\cite{DBLP:journals/ijon/SuALPBL24}. The intermediate representations $H$ are then fed into the FSQ module for quantization, and the quantized representations are passed through the rest of MinMo modules, including $\mathrm{Voice\ Encoder}_2$ and $\mathrm{MinMo\ LLM}$, to predict the posterior probabilities of the corresponding text tokens.

In the FSQ module, the intermediate representations $H$ are first projected into a $D$-dimensional low-rank space, and the values of each dimension are quantized into $[-K,K]$ with the bounded round operation $\mathrm{ROUND}$. Then, the quantized low-rank representations $\bar{H}$ are projected into the original dimension $\tilde{H}$, as follows:
\begin{equation}
\begin{split}
	\bar{H} &= \mathrm{ROUND}(\mathrm{Proj}_{down}(H)) \\
	\hat{H} &= \mathrm{Proj}_{up}(\bar{H})
\end{split}
\end{equation}
During the training stage, the straight-through estimation is used to approximate the gradients of the FSQ module and $\mathrm{Voice\ Encoder}_1$.
The speech token $\mu_i$ is obtained by calculating the index of the quantized low-rank representation $\bar{h}_i$ in the $(2K+1)$-ary system:
\begin{equation}\label{eq:codec}
	\mu_i = \sum_{j=0}^{D-1}{\bar{h}_{i,j}(2K+1)^{j}}
\end{equation}
Together the $\mathrm{Voice\ Encoder}_1$, the low-rank projector of the FSQ module, the bounded round operation, and the index calculation form the speech tokenizer of CosyVoice 3. Our speech tokenizer works at a token rate of 25 Hz, i.e., 25 speech tokens per second.

\subsection{Reinforcement Learning with Differentiable Reward Optimization}

Recent TTS systems~\cite{DBLP:journals/corr/abs-2406-02430,sun2025f5r} have demonstrated that reinforcement learning (RL) is effective in enhancing the quality of generated speech. However, to the best of our knowledge, a generally applicable RL methodology for speech generation has not been established. Unlike LLMs in the NLP task, TTS systems require additional downstream conditional flow matching (CFM) and vocoder models to convert discrete speech tokens into audio waveforms. The computational demands posed by these downstream models are substantial. More seriously, after downstream processing, the resulting voices consistently exhibit high similarity; therefore, it is challenging to differentiate between positive and negative feedback for training the reward model.

In order to address these issues, we introduce the \textbf{Differentiable Reward Optimization (DiffRO)} approach to directly optimize the speech tokens rather than the synthesized audio. DiffRO first trains an ASR-like Token2Text model on the ASR training data, then uses the posterior probability as the reward. To further simplify the training strategy, DiffRO uses the Gumbel-Softmax operation to sample the LLM predicted tokens and then directly optimize the speech tokens to maximize the reward score with back-propagation rather than the RL training loop:

\begin{equation}
\tilde{\mu}_t = \text{GumbelSoftmax} P_{\pi_{\theta}}(\mu_t | \mu_{1:t-1}; Y)   \label{eq:gumbel}
\end{equation} 
\begin{equation}
    R_{ASR}(Y) = \log P_{ASR}(\tilde{Y}_n = Y_n | Y_{1:n-1}; \tilde{\mu}_{1:T})   
\end{equation}
where $\mu_t$ and $\tilde{\mu}_t$ denote the ground-truth speech token and its sampled prediction at timestep $t$. $R_{ASR}$ is the reward function computed based on the ASR-like Token2Text model.
Since $R_{ASR}(Y)$ aims at encouraging $\tilde{\mu}$ to catch all information from the text, it can help the TTS system to comprehend the text clearly and accurately. Therefore, we can directly optimize the LLM to align the output tokens with ASR preference and use the Kullback-Leibler (KL) divergence to prevent the model from deviating too far from the reference model. Different from other RL methods, we compute the KL divergence on the \textbf{output token-level logits} rather than on the sequence-level posterior probability.

\begin{equation} \label{rl}
    \pi_\theta^* = \max_{\pi_\theta} \mathbb{E} \left[ R(Y) \right] - \beta D_{\text{KL}} \left[ \pi_\theta(\mu|Y) \| \pi_{\text{ref}}(\mu|Y) \right]
\end{equation}

\begin{equation}
    D_{\text{KL}} \left[ \pi_\theta(\mu|Y) \| \pi_{\text{ref}}(\mu|Y) \right] = \sum_{t=1}^{T}\sum_{k=0}^{Q} P_{\pi_{\theta}}(\mu_t=k) \log \left( \frac{P_{\pi_{\theta}}(\mu_t=k)}{P_{\pi_{\text{ref}}}(\mu_t=k)} \right)
\end{equation}
where $Q$ is the codecbook size of the FSQ module and equals to $(2K+1)^{D-1}$.

Besides the Token2Text model, DiffRO also uses other downstream tasks such as SER, MOS score prediction, AED, and other audio understanding tasks for \textbf{multi-task reward (MTR)} modeling. The MTR mechanism can help TTS systems to control the voice attributes $\{A_i\}_{i=1}^K$ by following instructions. 

\begin{equation}
R_{MTR}(Y, \{A_i\}_{i=1}^K) = \sum_i \log P_{\text{task}_i}(\tilde{A_i} = A_i | \tilde{\mu})
\end{equation}

\subsection{Pronunciation Inpainting}
LLM-based TTS systems predominantly use the BPE text tokenizer, taking raw text as input. Compared to traditional phoneme-based methods, these systems lack controllability in pronunciation. Specifically, when it comes to mispronunciations caused by polyphonic character or rare words that are sparse or do not appear in the training data, there lack robust methods that are based on human intervention.

To achieve an industry-level TTS system that is effectively controllable on pronunciations, we extend CosyVoice 3 to be able to model mixed sequences of words and phonemes with expansion of the vocabulary of tokenizer. To achieve this goal, we construct an auxiliary training set by replacing Chinese \textbf{monophonic} characters with pinyin and replacing English \textbf{monophonic} words with phonemes using the CMU pronunciation dictionary. This auxiliary dataset is added to the base training set.

\subsection{Self-training for Text Normalization}
Before text tokenization, TTS systems generally process the raw text by a text normalizaiton (TN) module to convert numbers and special symbols into their verbalization text, which relies on large amounts of hand-crafted rules; however, hand-crafted rules are constantly challenged by coverage on special symbols.

We explore LLMs for conducting the TN task, hence building a more unified end-to-end TTS system. Taking raw text as input,  we utilize three ways to construct another auxiliary training set: 1) We pass raw text through an internal rule-based TN module,  obtain text-normalized text, and synthesize audio by CosyVoice 2. 2) We prompt Qwen-Max\cite{Yang2024qwen2} to conduct text normalization and then synthesize audio on the normalized text by CosyVoice 2. 3) We prompt Qwen-Max to conduct inverse text normalization on text in existing text-audio pairs and obtain the raw text (that is, unnormalized text). The raw text and their corresponding audio are considered as a paired sample and directly added to the base traning set. We verify that the new system trained on the extended training set can synthesize raw text directly and exhibits better robustness and coverage on various special symbols. 

\subsection{Instructed Speech Generation}
To enhance controllability and expressiveness of CosyVoice 3, compared to CosyVoice 2, we integrate more expressive speech data into the base
training set. The duration of high-quality instruction-following data is expanded from 1,500 hours to 5,000 hours, covering a wider range of types including emotions, speed, voice tones, dialects, accents, and role-playing. The total number of types is increased to over 100, as illustrated in Table~\ref{tab:instruction-list}.
Similar to CosyVoice 2, CosyVoice 3 also supports language instructions and fine-grained instructions.
For natural language instructions, a natural language description and a special end token, ``\textless$|$endofprompt$|$\textgreater", is prepended to the input text for speech synthesis. 
For fine-grained instructions, vocal bursts between text tokens and vocal feature tags are supported for control. For example, markers such as ``[laughter]'' and ``[breath]'' in the input text can be used to generate a noticeable laughter and breath, respectively. The tag ``\textless strong\textgreater XXX\textless/strong\textgreater'' is used to indicate emphasis on specific words.
\begin{table}[thb]
\centering
\scalebox{1.0}{
\begin{tabular}{cccc}
\toprule
adventurous & ambitious & ancient & angry \\ 
artistic & authoritative & bold & brave \\ 
calm & charming & cheerful & clever \\ 
commanding & compassionate & confident & conflicted \\ 
contempt & courageous & creative & cunning \\ 
curious & dark & deceptive & dedicated \\ 
defiant & determined & disciplined & disgusted \\ 
empathetic & energetic & fearful & fearless \\ 
happy & heroic & hopeful & humble \\ 
imaginative & indifferent & insightful & intelligent \\ 
introspective & joyful & loyal & merciless \\ 
mysterious & noble & objective & optimistic \\ 
passionate & patient & proud & relaxed \\ 
relentless & responsible & sad & selfless \\ 
serious & shocked & stealthy & surprised \\ 
vengeful & vigilant & wise & fast \\ 
loud & slow & soft & adventurer \\ 
alchemist & architect & chef & craftsman \\ 
detective & doctor & girl & knight \\ 
leader & merchant & peppa & poet \\ 
robot & ruler & scholar & wanderer \\ 
warrior & witch & youth & anhui dialect \\ 
cantonese dialect & chongqing dialect & hebei dialect & shandong dialect \\ 
shanghai dialect & sichuan dialect & tianjin dialect & xi'an dialect \\ 
zhengzhou dialect & chinese english accent & indian english accent & russian english accent \\
\bottomrule
\end{tabular}
}
\vspace{0.1cm}
\caption{The 100 top-appeared speaking styles in pre-training data.}
\label{tab:instruction-list}
\end{table}


\subsection{Capability Transfer in Speaker Fine-tuning}
\subsubsection{Turning a Monolingual Speaker into a Polyglot}
\label{sec:polyglot}
A notable improvement in CosyVoice 3 over its predecessor is the extended language support. To enable a monolingual target speaker to speak multiple languages, we build an auxiliary training dataset, which contains studio-quality monolingual data from randomly-selected speakers covering all supported languages. The speaker ID and the language ID of every utterance are specified in a natural language instruction.
\begin{table}[thb]
\centering
\scalebox{1.0}{
\begin{tabular}{lr}
\toprule
\textbf{Examples} \\
\midrule
 - \begin{CJK}{UTF8}{gbsn}你是说话人小明。请讲法语。\end{CJK}         \\
 - You are Speaker B. Please speak German.         \\
\bottomrule
\end{tabular}
}
\vspace{0.1cm}
\caption{Examples of natural language instructions in the multilingual SFT dataset.}
\label{tab:mtlsftdata}
\end{table}

\subsubsection{Transferring the Capability of Instructed Generation}
Fine-tuning the pre-trained model with speaker-specific data can enhance the quality and expressiveness of generated output for individual speakers. We develop a training dataset that is partially labeled with speaker IDs. It includes high-quality data from the target speaker along with pre-training instruction-following dataset. In the natural language instruction prompt, we specify the speaker prompt and the style prompt. For example, a complete instruction prompt might be, ``You are Speaker A. Please talk to me happily.'' However, some data entries might lack speaker IDs or style labels; in such cases, we leave those fields blank in the prompt. During the fine-tuning process, we also randomly mask the speaker prompt or the style prompt to enhance the model's transfer capability. This method ensures comprehensive instruction coverage across different speakers and helps prevent potential catastrophic forgetting in instructed generation with the pretrained models.

\section{The Multilingual Data Pipeline}
Compared to Chinese and English, it is more challenging to acquire large-scale high-quality TTS data in other languages. To tackle this challenge, we collect in-the-wild multilingual audio data mainly from Internet audiobooks, videos, and podcasts. Then, we implement a multilingual data processing pipeline to produce model training data with sufficient quality. The pipeline consists of six steps, as follows: 1. Speech detection and segmentation; 2. Noise reduction; 3. ASR transcription; 4. Punctuation adjustment; 5. Volume standardization; and 6. Filtering out data with abnormal audio-text length ratios.

\paragraph{Speech detection and segmentation.}
Raw data is sequentially processed by speaker diarization, voice activity detection (VAD), and audio event detection modules. As a result, speaker-level speech segments shorter than 30 seconds are obtained. Although we use in-house modules in this step, they can be replaced by open-source alternatives to the same effect.

\paragraph{Noise reduction.}
We employ a MossFormer2~\cite{zhao2024mossformer2} model for noise reduction. Next, based on the energy levels of the leading and trailing frames of the utterances, ones starting or ending with incomplete words due to abnormal truncation are screened out; the remaining utterances, with leading and trailing silence trimmed, are retained for further processing.

\paragraph{ASR transcription.}
To obtain text transcriptions with adequate reliability, we first use Faster-Whisper Large-V3~\cite{systran_fwhisper_large_v3} for language identification, then employ different open-source ASR models, namely, Faster-Whisper Large-V3, NVIDIA NeMo Canary-1B~\cite{nvidia_canary_1b}, Meta FAIR seamlessM4T-V2-large~\cite{meta_seamlessm4t_v2}), to transcribe the utterances. We then perform cross validation and select transcriptions with an average pair-wise WER lower than 15\% among the ASR results from different systems.  

\paragraph{Punctuation adjustment.}
Since the punctuations in the ASR-generated texts may fail to properly represent the actual pauses in the corresponding audio, we use Montreal Forced Aligner \cite{mcauliffe17_interspeech} to derive the durations between words and clauses or phrases, then add or remove punctuations by preset thresholds ($\geq$300 milliseconds to add a comma while $\leq$50 milliseconds to remove punctuations indicating pauses, i.e. commas, semicolons, colons, full stops, question marks and exclamation marks).

\paragraph{Volume standardization.}
A simple and straightforward normalization is applied for volume standardization:
\begin{equation}
\text{normalized\_wav} = \frac{\text{raw\_wav}}{\text{max(raw\_wav)}} \times 0.6
\end{equation}

\paragraph{Filtering out utterances with abnormal audio-text length ratios.}
After all the above processing steps, speech tokens and text tokens are extracted for every generated utterance-text pair. Then, the utterance-level ratios of the lengths of the speech tokens and text tokens are calculated and sorted. We discard the utterances in the smallest 1\% and utterances in the largest 5\% in terms of the length ratios, to filter out possible abnormal cases with issues such as a short audio containing no human speech but corresponding to a long text transcription, or a long audio clip containing only a short human speech segment in the target language thus corresponding to a short text transcription.

\section{Experimental Settings}
\subsection{Training Data for Speech Tokenizer}

A 530,000-hour supervised multi-task dataset is used to train the speech tokenizer with normalized transcriptions as labels, including automatic speech recognition (ASR), language identification (LID), speech emotion recognition (SER), audio event detection (AED), and speaker analysis (SA). Details of the training data are listed in Table~\ref{tab:fsqdata}. 
The multilingual ASR training data consists of Chinese, English, Japanese, Korean, Russian, French, and German.

\begin{table}[thb]
\centering
\scalebox{1.0}{
\begin{tabular}{lr}
\toprule
\textbf{Language} & \textbf{Duration (hours)} \\ \midrule
Automatic Speech Recognition (ASR) - Multilingual  & 365K          \\
Language Identification (LID)  & 85K           \\ 
Speech Emotion Recognition (SER)  & 48K           \\ 
Audio Event Detection (AED) & 21K           \\ 
Speaker analysis & 11K           \\ 

\bottomrule
\end{tabular}
}
\vspace{0.1cm}
\caption{Details of the training data for speech tokenizer.}
\label{tab:fsqdata}
\end{table}

\subsection{Scaling up Dataset Size and Model Size for CosyVoice 3}
\begin{figure}[!t]
\centering
\subfloat[Minority languages]{\includegraphics[width=0.48\linewidth]{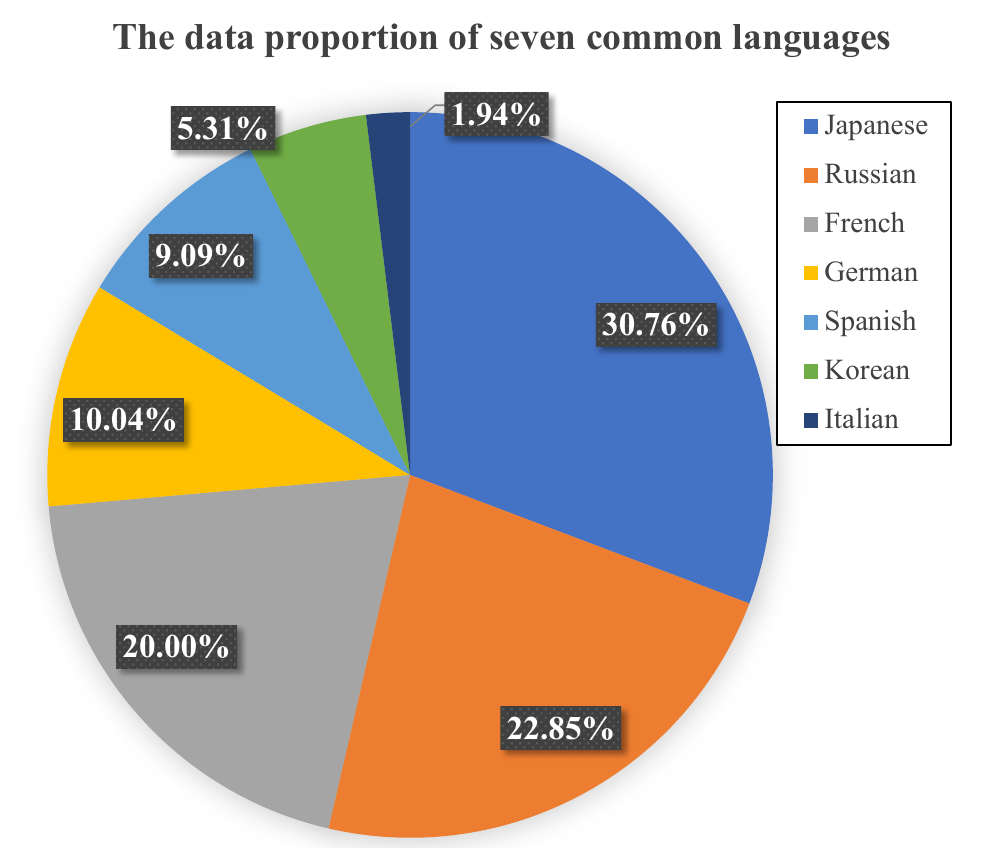}
	\label{fig:mino-lang}}
\subfloat[Chinese dialects]{\includegraphics[width=0.48\linewidth]{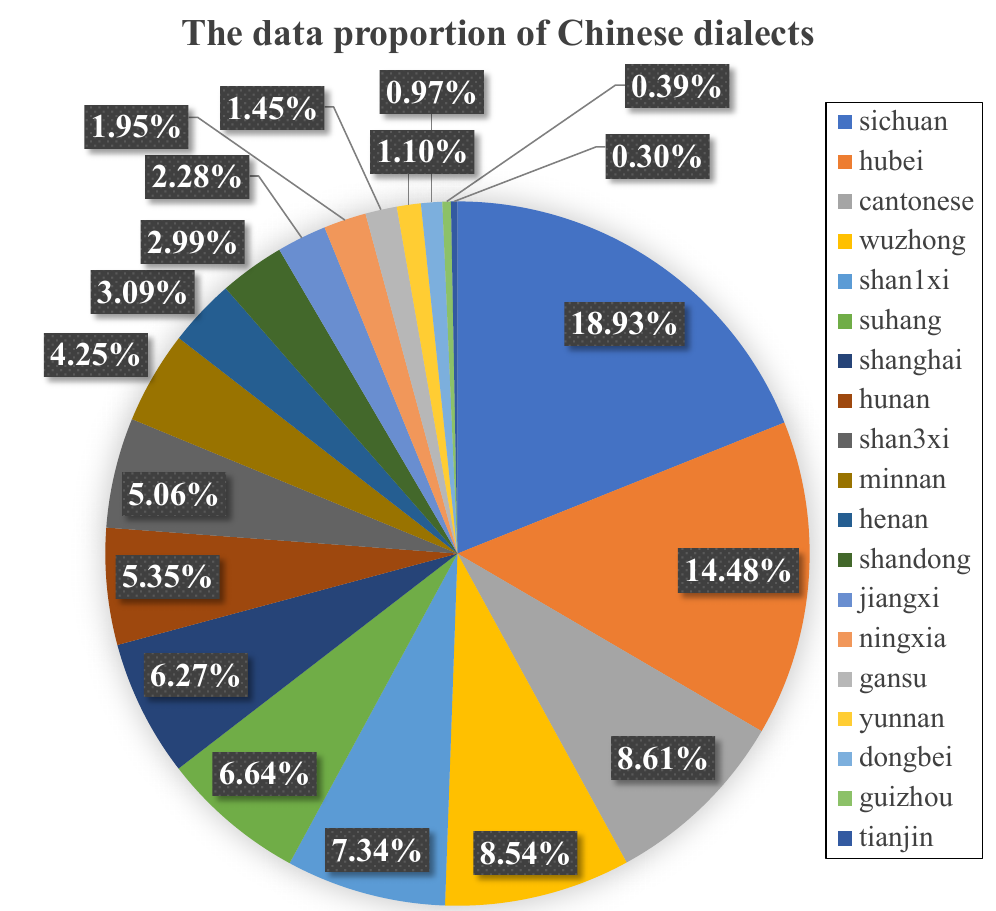}
	\label{fig:cn-dialect}}
\caption{The data percentage of (a) seven minority languages and (b) 19 Chinese accents or dialects.}
\label{fig:proportion}
\end{figure}

In CosyVoice 3, we scale up the data volume from multiple aspects. For widely used Chinese and English data, we employ a combination of low-cost data production pipelines and self-training data construction to enhance diversity in domains, styles, text formats, and rare cases.
Regarding domain diversity, we collect voice data from various fields such as e-commerce, navigation, finance, and education. In terms of style diversity, we add conversations, speeches, singing, and more. For text diversity, we construct different text formats for the same speech through text normalization (TN) and inverse text normalization (ITN), enhancing the model's robustness to varied text formats. Additionally, we use self-training to strategically create numerous rare cases with an early version of CosyVoice 3 to improve synthesis stability. In terms of language coverage, we augment the Chinese and English dataset with seven common languages, including Japanese, Russian, French, German, Spanish, Korean, and Italian, with the data percentage shown in Figure~\ref{fig:mino-lang}. Our previous work~\cite{cosyvoice} shows that the supervised multi-task speech tokenizer could performs well on some unseen languages (that is, Spanish and Italian in the case of CosyVoice 3).
In addition to standard common tongue pronunciations, we increase the coverage of Chinese accents and dialects, supporting 19 common accents or dialects, with the data percentage shown in Figure~\ref{fig:cn-dialect}. Through these data scaling efforts, the training data of CosyVoice 3 reaches one million hours, covering the majority of user cases in daily life and advancing towards in-the-wild zero-shot speech generation.

In addition to scaling dataset size, scaling up model size is crucial for current large-scale models. Therefore, we increase the model sizes of both the text-to-speech language model (LM) and the Conditional Flow Matching (CFM) model in CosyVoice 3. Specifically, the text-to-speech LM is increased from 0.5B to 1.5B parameters. For the CFM, we adopt the recent diffusion transformer (DiT) \cite{peebles2023scalable,DBLP:journals/corr/abs-2410-06885} as the backbone, increasing the number of parameters from 100M to 300M. Preliminary experiments demonstrate the strong performance of the DiT architecture; hence, the complicated text encoder and the length regularization module are no longer needed and removed from CosyVoice 3. We solve the frame rate mismatch issue between speech tokens and Mel features by a simple interpolation operation.

\subsection{Evaluation Settings for Zero-shot Capability}
For evaluating CosyVoice 3's zero-shot speech generation capabilities, we focus on three key aspects: content consistency, speaker similarity, and audio quality. For content consistency, we measure the Character Error Rate (CER) or Word Error Rate (WER) of the ASR transcription against the given text, using Whisper-large V3 \cite{DBLP:conf/icml/RadfordKXBMS23} for English ASR and Paraformer \cite{gao2023funasr} for Chinese ASR. To assess speaker similarity, we extract speaker embeddings from the generated speech using the ERes2Net speaker verification model~\cite{chen2023enhanced} and calculate the cosine similarity with the embedding of the reference speech. For audio quality, we score the generated speech using the DNSMOS network~\cite{DBLP:conf/icassp/ReddyGC22}, the scores of which show high correlations with human auditory perception.

We conduct evaluations on two test sets. The first is the widely used SEED-TTS-Eval test set~\cite{DBLP:journals/corr/abs-2406-02430}, where test cases are categorized into Mandarin, English, and hard Chinese subsets. To facilitate fair comparison with other models, we also use a WavLM-based speaker recognition model to calculate the speaker similarity~\cite{chen2022large}. Notably, recent advances in speech generation models have left little room for improvements, with models achieving quite similar scores; hence,
we introduce a new multilingual benchmark \textbf{CV3-Eval} for evaluation, detailed in Section \ref{sec:mult-ling-bench}.

To perform a comprehensive comparison with CosyVoice 3, we employ 10 commonly-used speech generation models as the baselines, which achieve state-of-the-art (SOTA) or competitive performance in some aspects. Specifically, non-autoregressive (NAR) models include MaskGCT \cite{DBLP:journals/corr/abs-2409-00750}, E2 TTS \cite{DBLP:journals/corr/abs-2406-18009}, F5-TTS \cite{DBLP:journals/corr/abs-2410-06885}, and F5R-TTS \cite{sun2025f5r}, while autoregressive (AR) baselines are Seed-TTS \cite{DBLP:journals/corr/abs-2406-02430}, FireRedTTS \cite{DBLP:journals/corr/abs-2409-03283}, Qwen2.5-Omni \cite{xu2025qwen2}, CosyVoice \cite{cosyvoice}, CosyVoice 2 \cite{du2024cosyvoice2}, and Spark TTS \cite{wang2025spark}.

\subsection{CV3-Eval: a Multilingual Benchmark}
\label{sec:mult-ling-bench}
With the rapid development of speech generation models, existing evaluation benchmarks no longer meet the model assessment requirements, especially for zero-shot voice cloning. Firstly, most evaluation benchmarks such as Librispeech~\cite{librispeech} are sampled from audio books, where the speaker's pronunciations are clean and standard. As a result, some systems can effortlessly synthesize high-quality audio that even beats the ground truth audio. However, source audio is often noisy in real-world application scenarios, presenting challenges that these benchmarks fail to address. Secondly, most benchmarks are designed for Chinese and English, while multilingual evaluation benchmarks are absent. Finally, traditional benchmarks only focus on the pronunciation accuracy, speaker similarity, and the MOS scores for audio quality. These evaluation metrics cannot accurately measure the comprehensive capability of a TTS system, including aspects such as emotion expression, rhythmic richness, voice controllability, and cross-lingual voice cloning. 

To better evaluate CosyVoice 3, we establish a multilingual benchmark, CV3-Eval, including subsets for both objective and subjective evaluation.  

\paragraph{Objective Evaluation.}
The objective evaluation subset is further split into three subsets, including multilingual voice cloning, cross-lingual voice cloning, and emotion cloning, as follows:
\begin{itemize}[leftmargin=*]
\item \textbf{Multilingual Voice Cloning}: The multilingual voice cloning subset contains 9 languages with 500 samples for each language, including Chinese (zh), English (en), Japanese (ja), Korean (ko), German (de), France (fr), Russian (ru), Italian (it), and Spanish (es). The source audio and target text are sampled from CommonVoice~\cite{commonvoice} and FLUERS~\cite{fleurs} datasets. To simulate real-world application scenarios, we do not filter out audio with noisy background or long silence, which poses challenges to the robustness of the TTS system. In addition, we construct two hard-case test sets for Chinese and English, where the target text includes rare words, tongue twisters, domain-specific terms, etc.
\item \textbf{Cross-lingual Voice Cloning}. For the cross-lingual voice cloning subsets, the source audio and target text are from different languages, including zh, en, ja, and ko. This subset can evaluate the language transfer capability of the TTS system.
\item \textbf{Emotion Cloning}. The audio prompts in the emotion cloning subset are sourced from EmoBox\cite{emobox} and SeCap\cite{xu2023secap}, including both Chinese and English samples. Due to the insufficient expressiveness of some emotion labels, we only include samples labeled as happy, sad, or angry, with 100 samples for each language. We further categorize these samples into text-related and text-unrelated parts, depending on whether the target text is semantically consistent with the target emotion. This helps us determine whether the synthetic emotional features are primarily derived from the text content or the prompt audio.

\end{itemize}

\paragraph{Subjective Evaluation.}
Besides the objective evaluation subset, we also prepare three subjective subsets for expressive voice cloning, expressive voice continuation, and Chinese accent voice cloning.

\begin{itemize}[leftmargin=*]
\item \textbf{Expressive Voice Cloning}. To explore the model’s capacity for generating expressive speech, the Expressive Voice Cloning benchmark is designed to include audio prompts with distinctive features, such as highly emotional intonation, whisper and shout, and extreme slow or fast speaking rate. Audio prompts are selected from different challenging application scenarios such as news, podcasts, TV drama, academic reports, poetry recitation, etc. Voices of some public figures are also sampled for evaluation. 

\item \textbf{Expressive Voice Continuation}. 
Due to the high variability in human perception, achieving a fair subjective evaluation of expressive voice cloning is challenging. To mitigate this issue, we design a voice continuation task. Specifically, we select 120 audio samples with different emotions, rhythms, speeds, and volumes from the website and cut the first 3 seconds of the audio clip as prompt speech.
Therefore, we can evaluate the synthesized remaining speech based on its similarity with the ground truth speech.

\item \textbf{Chinese Accent Voice Cloning}. Since there is currently no reliable objective method to evaluate the authenticity of accents, we construct a subjective evaluation dataset for Chinese dialects. The dataset includes 18 different Chinese dialects, such as Cantonese, Dongbei, Minnan, Shanghai dialects, etc. All prompt speech samples are sourced from in-house industrial data.

\end{itemize}


\section{Experimental Results}

\subsection{Objective TTS Results on SEED-TTS-Eval}

\begin{table*}[htb]
	\centering
	\setlength\tabcolsep{4.5pt}
	\scalebox{0.80}{
		\begin{tabular}{lclclcl}
			\toprule
			\multirow{2}{*}{\textbf{Model}} & \multicolumn{2}{c}{\textbf{\emph{test-zh}}} & \multicolumn{2}{c}{\textbf{\emph{test-en}}} & \multicolumn{2}{c}{\textbf{\emph{test-hard}}} \\
			\cmidrule(r){2-3} \cmidrule(r){4-5} \cmidrule(r){6-7}
			& \textbf{CER (\%)~$\downarrow$} & \multicolumn{1}{c}{\textbf{SS~$\uparrow$}} & \textbf{WER (\%)~$\downarrow$} & \multicolumn{1}{c}{\textbf{SS~$\uparrow$}} & \textbf{CER (\%)~$\downarrow$} & \multicolumn{1}{c}{\textbf{SS~$\uparrow$}}  \\
			\midrule
			\textbf{Human} & 1.26 & 0.755~(0.775) & 2.14 & 0.734~(0.742)  & - & \multicolumn{1}{c}{-} \\
			\textbf{Vocoder Resyn.} & 1.27 & 0.720 & 2.17 & 0.700 & - & \multicolumn{1}{c}{-} \\
			\midrule
                \multicolumn{7}{c}{\textbf{Non-autoregressive Models}} \\
                \midrule
			\textbf{MaskGCT}~\cite{DBLP:journals/corr/abs-2409-00750} & 2.27 & 0.774~(0.752) & 2.62 & 0.714~(0.730)  & 10.27 & 0.748~(0.720) \\
			\textbf{E2 TTS (32 NFE)}~\cite{DBLP:journals/corr/abs-2406-18009} & 1.97 & 0.730 & 2.19 & 0.710  & - & \multicolumn{1}{c}{-} \\
			\textbf{F5-TTS (32 NFE)}~\cite{DBLP:journals/corr/abs-2410-06885} & 1.56 & 0.741~(0.794) & 1.83 & 0.647~(0.742)  & 8.67 & 0.713~(0.762) \\
                \textbf{F5R-TTS}~\cite{sun2025f5r} & 1.37 & 0.754 & - & \multicolumn{1}{c}{-}  & 8.79 & 0.718 \\
                \midrule
                \multicolumn{7}{c}{\textbf{Autoregressive Models}} \\
                \midrule
                \textbf{Seed-TTS}~\cite{DBLP:journals/corr/abs-2406-02430} & {1.12} & \textbf{0.796} & 2.25 & \textbf{0.762}  & 7.59 & \textbf{0.776} \\
			\textbf{FireRedTTS}~\cite{DBLP:journals/corr/abs-2409-03283} & 1.51 & 0.635~(0.653)  & 3.82 & 0.460~(0.526)  & 17.45 & 0.621~(0.639) \\
                \textbf{Qwen2.5-Omni-7B}~\cite{xu2025qwen2} & 1.70 & 0.752 & 2.72 & 0.632  & 7.97 & 0.747 \\
                \textbf{Qwen2.5-Omni-7B$_{RL}$}~\cite{xu2025qwen2} & 1.42 & 0.754 & 2.33 & 0.641  & 6.54 & 0.752 \\
                \textbf{CosyVoice} \cite{cosyvoice} & 3.63 & 0.723~(0.775) & 4.29 & 0.609~(0.699)  & 11.75 & 0.709~(0.755) \\
			\textbf{CosyVoice 2}\cite{du2024cosyvoice2} & 1.45 & 0.748~(0.806) & 2.57 & 0.652~(0.736)  &  6.83 & 0.724~(0.776) \\
                \textbf{Spark TTS}\cite{wang2025spark} & 1.20 & 0.672 & {1.98} & 0.584  & - & \multicolumn{1}{c}{-} \\
                \midrule
                \textbf{CosyVoice 3-0.5B} & 1.16 & 0.780~(\textbf{0.840}) & 2.02 & 0.718~(\textbf{0.790}) & 6.08 & \underline{0.758}~(\underline{0.815}) \\
                \textbf{CosyVoice 3-0.5B$_{{RL}}$} & \underline{0.75} & 0.774~({0.836}) & \underline{1.76} & 0.695~(0.783) & \textbf{5.09} & 0.750~(0.809) \\
                \textbf{CosyVoice 3-1.5B} & {1.12} & \underline{0.781}~(\underline{0.837}) & 2.21 & \underline{0.720}~(\underline{0.789}) & 5.83 & \underline{0.758}~(\textbf{0.816}) \\
                \textbf{CosyVoice 3-1.5B$_{{RL}}$} & \textbf{0.71} & 0.775~(0.836) & \textbf{1.45} & 0.695~(0.784) & \underline{5.66} & 0.750~(0.810) \\
			\bottomrule
	\end{tabular}}
	\caption{Zero-shot TTS performance comparison between CosyVoice 3 and the baselines on the SEED test sets in terms of content consistency (WER/CER) and speaker similarity (SS). For speaker similarity, the results outside parentheses are measured by WavLM-based models while the results inside parentheses are measured by ERes2Net. While the \textbf{boldface} denotes the best result, the \underline{underline} denotes the second best.}
	\label{tab:tts-seed-test}
\end{table*}

Table~\ref{tab:tts-seed-test} presents the TTS performance of CosyVoice 3 and several recent models across the SEED test sets, which include the Chinese \emph{test-zh}, English \emph{test-en}, and the challenging \emph{test-hard} sets. The evaluation focuses on content consistency (WER/CER) and speaker similarity (SS).

For \textbf{content consistency}, CosyVoice 3 achieves significant improvements over CosyVoice 2, with relative gains of 44\% on \emph{test-zh} and 51\% on \emph{test-en}. In the \emph{test-hard} set, CosyVoice 3 reduces the CER from 6.83\% to 5.09\% (26\% relative improvement). Compared to other baselines, CosyVoice 3 consistently excels across all metrics. Notably, CosyVoice 3-1.5B$_{RL}$ records the lowest CER of 0.71\% in \emph{test-zh} and the lowest WER of 1.45\% in \emph{test-en}, showcasing its superior synthesis accuracy. In the challenging \emph{test-hard} scenario, CosyVoice 3-0.5B$_{RL}$ achieves the lowest CER of 5.09\%, while the 1.5B variant follows closely with 5.66\%. The larger model's underperformance compared to the smaller one is due to the limited dataset available for pretraining and post-training, particularly in challenging scenarios. We plan to expand our dataset to tens of millions of hours to support more effective training of larger models in the future.

Regarding \textbf{speaker similarity}, CosyVoice 3 demonstrates a strong ability to replicate speaker characteristics accurately. It outperforms CosyVoice 2 and other baselines, except Seed-TTS, as shown through both WavLM-based and ERes2Net measurements. The similarity gap between CosyVoice 3 and Seed-TTS is primarily due to differences in speaker diversity and pretraining data volume. Enhancing speaker similarity in CosyVoice 3 can be achieved by scaling up pretraining data, a direction we intend to pursue in future work. Additionally, Table~\ref{tab:tts-seed-test} shows that RL post-training contributes to 12\% to 35\% relative improvements in content consistency, enhancing robustness and adaptability in multilingual and complex synthesis tasks. With RL post-training, CosyVoice 3 establishes a new state of the art in TTS performance, demonstrating substantial advancements over previous models.

\subsection{Objective Evaluation on Multilingual Benchmark CV3-Eval}
\subsubsection{Results of Multilingual Voice Cloning}

We evaluate CosyVoice 3 against competitive open-source TTS systems, including F5-TTS, Spark-TTS, and GPT-SoVits\footnote{https://github.com/RVC-Boss/GPT-SoVITS}, using the Multilingual Voice Cloning subset of CV3-Eval benchmark. Table~\ref{tab:ml-clone} provides CERs for Chinese, Japanese, and Korean, and WERs for English, German, Spanish, French, Italian, and Russian. The Multi-lingual Voice Cloning subset proved to be significantly challenging, as CosyVoice 3 is the only system capable of covering all languages in this subset. For most languages, the performance difference between CosyVoice3-0.5B and CosyVoice3-1.5B is minimal. Furthermore, as shown in Table \ref{tab:hard-ml-clone}, generating rare words, tongue twisters, and domain-specific terms remains difficult for CosyVoice 3, highlighting areas for future improvement.


\begin{table}[t]
\centering{%
\begin{tabular}{l|ccccccccccc}
\toprule
\textbf{Model} & \textbf{zh} & \textbf{en} & \textbf{ja} & \textbf{ko} & \textbf{de} & \textbf{es} & \textbf{fr} & \textbf{it} & \textbf{ru} \\ 
\midrule
\textbf{F5-TTS} & 5.47 & 8.90 & -- & -- & -- & -- & -- & -- & --  \\
\textbf{Spark-TTS} & 5.15 & 11.0 & -- & -- & -- & -- & -- & -- & --  \\
\textbf{GPT-SoVits} & 7.34 & 12.5 & -- & -- & -- & -- & -- & -- & --  \\
\midrule
\textbf{CosyVoice2} & 4.08 & 6.32 & 9.13 & 19.7 & -- & -- & -- & -- & --  \\
\textbf{+ DiffRO} &   3.00 & 4.72 & 6.36 & 5.14  & -- & -- & -- & -- & --  \\
\textbf{CosyVoice3-0.5B} & 3.89 & 5.24 & 10.4 & 12.8 & 7.41 & 4.25 & 12.9 & 6.68 & 6.77 \\
\textbf{+ DiffRO}  & 2.89 & 3.68 & 5.15 & 4.02 & 4.51 & 2.99 & 8.56 & 2.94 & 3.79  \\
\textbf{CosyVoice3-1.5B} & 3.91 & 4.99 & 7.57 & 5.69 & 6.43 & 4.47 & 11.8 & 10.5 & 6.64  \\
\textbf{+ DiffRO} & 3.01 & 3.71 & 5.27 & 4.01 & 3.93 & 3.26 & 8.09 & 2.72 & 4.11 \\ 

\bottomrule
\end{tabular}
}
\vspace{0.05cm}
\caption{CER(\%) and WER(\%) on the CV3-Eval Multilingual Voice Cloning subset. -- means the language is unsupported.}
\label{tab:ml-clone}
\end{table}

\begin{table}[t]
\centering{%
\begin{tabular}{l|ccc|ccc}
\toprule
\multirow{2}{*}{\textbf{Model}} & \multicolumn{3}{c|}{\textbf{hard-zh}} & \multicolumn{3}{c}{\textbf{hard-en}} \\
 & \textbf{WER} & \textbf{SS} & \textbf{DNSMOS} & \textbf{WER} & \textbf{SS} & \textbf{DNSMOS} \\ 
\midrule
\textbf{CosyVoice2} & 12.58 & 72.6 & 3.81 & 11.96 & 66.7 & 3.95 \\
\textbf{+ DiffRO} & 10.66 & 71.7 & 3.81 & 10.25 & 62.4 & 3.97 \\
\textbf{CosyVoice3-0.5B}  & 14.15 & 78.6 & 3.75 & 9.04 & 75.9 & 3.92 \\
\textbf{+ DiffRO}  & 8.26 & 77.8 & 3.80 & 7.60 & 73.9 & 3.95 \\
\textbf{CosyVoice3-1.5B} & 9.77 & 78.5 & 3.79 & 10.55 & 76.1 & 3.95 \\
\textbf{+ DiffRO} & 9.06 & 78.2 & 3.81 & 7.56 & 74.6 & 3.95 \\

\bottomrule
\end{tabular}
}
\vspace{0.05cm}
\caption{WER(\%), Speaker Similarity (SS), and MOS scores on the \textbf{hard samples} in the CV3-Eval Multilingual Voice Cloning subset.}
\label{tab:hard-ml-clone}
\end{table}

\subsubsection{Results of Cross-lingual Voice Cloning}



\begin{table}[t]
\centering{%
\scalebox{0.90}{
\begin{tabular}{l|ccc|ccc|ccc|ccc}
\toprule
\multirow{2}{*}{\textbf{Model}}  & \multicolumn{3}{c|}{\textbf{to-zh}} & \multicolumn{3}{c|}{\textbf{to-en}} & \multicolumn{3}{c|}{\textbf{to-ja}} & \multicolumn{3}{c}{\textbf{to-ko}} \\ 
  & \textbf{en} & \textbf{ja} & \textbf{ko} & \textbf{zh} & \textbf{ja} & \textbf{ko} & \textbf{zh} & \textbf{en} & \textbf{ko} & \textbf{zh} & \textbf{en} & \textbf{ja} \\
\midrule
\textbf{CosyVoice2} & 13.5 & 48.1 & 7.70 & 6.47 & 17.1 & 11.2 & 13.1 & 14.9 & 5.86 & 24.8 & 21.9 & 21.5 \\
\textbf{CosyVoice3-0.5B}  & 8.48 & 6.86 & 5.24 & 4.99 & 6.83 & 5.86 & 18.3 & 16.8 & 4.99 & 41.0 & 20.4 & 12.8 \\
\textbf{+ DiffRO}   & 5.16 & 3.22 & 1.03 & 3.40 & 4.41 & 4.78 & 7.91 & 7.25 & 3.29 & 16.9 & 11.6 & 8.2 \\
\textbf{CosyVoice3-1.5B} & 8.01 & 6.78 & 3.30 & 4.32 & 5.39 & 5.94 & 13.7 & 13.4 & 4.19 & 31.6 & 14.0 & 10.5 \\
\textbf{+ DiffRO} & 5.09 & 3.05 & 1.06 & 2.98 & 4.20 & 4.19 & 7.08 & 6.80 & 3.93 & 14.4 & 5.87 & 7.92 \\

\bottomrule
\end{tabular}}
}
\vspace{0.05cm}
\caption{WER(\%) results on the CV3-Eval Cross-lingual Voice Cloning subset.}
\label{tab:cl-clone}
\end{table}


\begin{table}[t]
\centering{%

\begin{tabular}{l|ccc|ccc}
\toprule
\multirow{2}{*}{\textbf{Model}}  & \multicolumn{3}{c|}{\textbf{en2zh}} & \multicolumn{3}{c}{\textbf{zh2en}}  \\ 
 & \textbf{WER} & \textbf{SS} & \textbf{MOS} & \textbf{WER} & \textbf{SS} & \textbf{MOS} \\ 
\midrule
\textbf{F5-TTS} & 11.6 &	64.2 & 3.77	& 5.57 & 64.7 & 3.77 \\
\textbf{Spark-TTS} & 12.4 & 48.4 & 3.65 & 7.36 & 56.7 &	3.61 \\
\midrule
\textbf{CosyVoice2} & 13.5 & 63.3 & 3.87 & 6.47 & 64.3 & 3.75 \\
\textbf{CosyVoice3-0.5B}  &  8.48 & 67.4 & 3.82 & 4.99 & 67.8 & 3.75 \\
\textbf{CosyVoice3-1.5B} &  8.01 & 66.9 & 3.83 & 4.32 & 66.4 & 3.77 \\

\bottomrule
\end{tabular}
}
\vspace{0.05cm}
\caption{WER(\%), Speaker Similarity (SS), and MOS scores from CosyVoice 3 and the baselines on the zh2en and en2zh voice cloning tasks.}
\label{tab:cl_zh2en}
\end{table}

Table~\ref{tab:cl-clone} illustrates the significant improvements CosyVoice 3 offers over CosyVoice 2 in cross-lingual voice cloning. Notably, CosyVoice 2 struggles with transferring voice from Japanese to Chinese due to the character overlap of two languages, a problem resolved in CosyVoice 3 by converting all Japanese characters into kana. Additionally, scaling the model size proves beneficial: CosyVoice3-1.5B exhibits better WERs across all conditions compared to CosyVoice3-0.5B, while maintaining similar speaker similarity. This indicates that larger models can enhance performance on challenging tasks due to increased capacity.

Since most open-source TTS systems only support Chinese and English, we further evaluate CosyVoice 3 against baselines for the zh2en and en2zh cross-lingual voice cloning tasks, as shown in Table~\ref{tab:cl_zh2en}. Compared to CosyVoice 3, F5-TTS and Spark-TTS show inferior performance on WER, with Spark-TTS also lagging significantly in SS compared to F5-TTS and CosyVoice 3. Regarding MOS scores, CosyVoice 3 demonstrates better results for en2zh and comparable results for zh2en. Overall, CosyVoice3-1.5B remains the leading model for zh2en and en2zh cross-lingual transfer tasks.


\subsubsection{Results of Emotional Voice Cloning}

\begin{table}[t!]
\centering{%
\begin{tabular}{l|ccc|ccc}
\toprule
\multirow{2}{*}{\textbf{Model}}  & \multicolumn{3}{c|}{\textbf{Text-Related}} & \multicolumn{3}{c}{\textbf{Text-Unrelated}} \\ 
  & \textbf{happy} & \textbf{sad} & \textbf{angry} & \textbf{happy} & \textbf{sad} & \textbf{angry}  \\ 
\midrule
\textbf{F5-TTS} & 0.92  & 0.52 & 0.72 & 0.80 & 0.28 & 0.64 \\
\textbf{Sparks-TTS} & 0.80  & 0.56 & 0.50 & 0.50 & 0.60 & 0.36 \\
\textbf{GPT-SoVits} & 0.88  & 0.54 & 0.50 & 0.48 & 0.40 & 0.30 \\
\midrule
\textbf{CosyVoice2} & 0.84  & 0.72 & 0.58 & 0.56 & 0.44 & 0.38 \\
\textbf{CosyVoice3-0.5B}  & 0.92 & 0.70 & 0.72 & 0.64 & 0.42 & 0.58 \\
\textbf{CosyVoice3-1.5B} & 0.86 & 0.64 & 0.72  & 0.64 & 0.44 & 0.48 \\
\textbf{+ DiffRO-EMO} & 0.98 & 0.68 & 0.84  & 0.98 & 0.50 & 0.68 \\

\bottomrule
\end{tabular}
}
\vspace{0.05cm}
\caption{Emotion Accuracy on the Text-Related and Text-Unrelated subsets of the CV3-Eval Emotional Voice Cloning subset.}
\label{tab:emo}
\end{table}

In the CV3-Eval Emotional Voice Cloning subset, we employe the emo2vec-large-plus model\footnote{https://www.modelscope.cn/models/iic/emotion2vec\_plus\_large/summary} as a classifier to assess the emotion expression capabilities of TTS systems. The results, displayed in Table~\ref{tab:emo}, reveal that most TTS systems perform well on text-related subsets, with CosyVoice 3 achieving the highest performance. Each system excels in expressing specific emotions, with "happy" being the easiest emotion to convey across all models. However, in text-unrelated tasks, emotion accuracy drops significantly, particularly for "sad" and "angry" emotions. This indicates that TTS systems primarily infer the emotional tone of output audio from text sentiment. This observation provides valuable insights into the less satisfactory performances and highlights areas for future improvement.


\subsubsection{Subjective Evaluation Results}
\begin{figure}[tbh]
    \centering
    \includegraphics[width=1.0\textwidth]{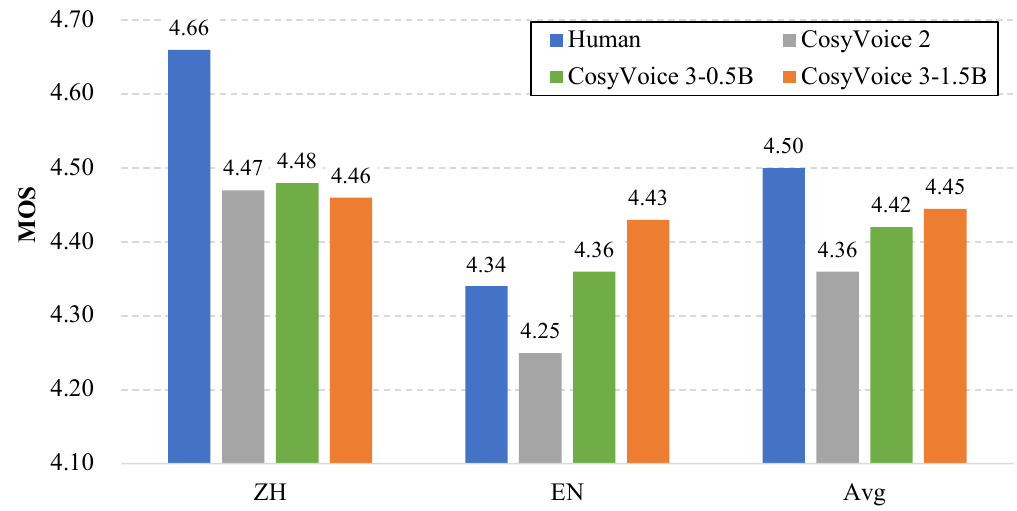}
    \caption{The Mean Opinion Scores (MOS) of zero-shot cloning models on Chinese, English and their average.}
    \label{fig:zero-shot-mos}
\end{figure}

In addition to objective metrics, we performe a subjective evaluation using Mean Opinion Scores (MOS). The test samples comprise 200 Chinese and English sentences, each assessed by 10 native speakers (5 male and 5 female). Scores ranged from 1 to 5, in 0.5-point increments. Figure \ref{fig:zero-shot-mos} shows the MOS scores for the CosyVoice 2, CosyVoice 3-0.5B, and CosyVoice 3-1.5B models across both languages, along with their average scores.

For Chinese, all three models perform similarly but still lag behind human speech. In English, CosyVoice 2 scores lower than human benchmarks, CosyVoice 3-0.5B matches human scores, and CosyVoice 3-1.5B scores notably higher. Overall, CosyVoice 3-1.5B outperforms CosyVoice 3-0.5B, with both surpassing CosyVoice 2, illustrating the advantages of data and model scaling.

Despite some differences from human speech in Chinese, CosyVoice 3 models still score above 4.45. This gap is primarily due to a few low-scoring cases in the synthetic output compared to real speech, indicating the need for improved synthesis stability in future work.

\subsection{Ablation of Speech Tokenizer}
\subsubsection{Up-streaming Recognition Tasks}
\begin{table}[htb]
\centering{%
\begin{tabular}{l|cccccc}
\toprule
\multirow{1}{*}{\textbf{Method}} & \textbf{C.V. EN} & \textbf{C.V. CN} & \textbf{C.V. JA} & \textbf{C.V. KO} & \textbf{Fluers EN} & \textbf{Fluers CN} \\ \midrule
\textbf{SenseVoice}   & 7.70 & 8.67 & - & - & 4.57 & 6.98 \\
\textbf{MinMo}   & 7.36 & 8.56 & - & - & 4.43 & 6.71 \\
\textbf{VQ-SenseVoice}   & 18.26 & 11.56 & - & - & 7.65 & 5.03 \\
\textbf{FSQ-SenseVoice} & 10.67 & 7.29 & - & - & 6.58 & 4.43 \\ 
\textbf{FSQ-MinMo} & 11.36 & 9.21 & 13.90 & 9.78  & 4.46 & 3.35 \\ 
\bottomrule
\end{tabular}
}
\vspace{0.05cm}
\caption{Comparison between VQ and FSQ inside the Sensevoice-large and MinMo encoders in terms of ASR WERs and CERs (\%) across language-specific subsets of the CommonVoice (\textbf{C.V.}) and the Fluers benchmarks. \textbf{FSQ-MinMo} is the tokenizer used in CosyVoice 3.}
\label{fsqres}
\end{table}

\begin{table}[htb]
\centering{%
\setlength\tabcolsep{5pt}
\begin{tabular}{l|cccccc}
\toprule
\multirow{1}{*}{\textbf{Method}} & \textbf{Language ID} & \textbf{Gender}  & \textbf{Age} & \textbf{Emotion} & \textbf{Vocal Sound } & \textbf{Sound Question} \\ \midrule
\textbf{MinMo}   & 99.2 & 84.8 & 70.1 & 62.4 & 90.7 & 59.1 \\
\textbf{FSQ-MinMo} & 99.2 & 72.8 & 41.8 & 68.4  & 61.3 & 57.7 \\ 
\bottomrule
\end{tabular}
}
\vspace{0.05cm}
\caption{Performance comparison between MinMo and FSQ-MinMo models in terms of Accuracy on the AIR-Bench benchmark, including Language ID, Gender, Age, Emotion, Vocal Sound, and Sound Question classification tasks.}
\label{tab:ml-asr}
\end{table}

Our supervised multi-task learning-based speech tokenizer exhibits strong performance across various speech and sound tasks. Specifically, as shown in Table~\ref{fsqres}, the FSQ-MinMo-based speech tokenizer in CosyVoice 3 effectively maintains multilingual ASR capabilities. By focusing exclusively on speech-related tasks in FSQ-MinMo and excluding others from the training set, we achieve superior recognition performance compared to MinMo on the Fluers CN test set. Additionally, Table~\ref{tab:ml-asr} illustrates that the FSQ-MinMo model performs comparably to the MinMo model on the AIR-Bench benchmark, which includes tasks such as LID, Gender, Age, Emotion, Vocal Sound, and Sound Question classification.


\subsubsection{Down-streaming TTS Tasks}
\begin{table*}[t!]
	\centering
	\setlength\tabcolsep{4.5pt}
	\scalebox{1.0}{
		\begin{tabular}{lclclcl}
			\toprule
			\multirow{2}{*}{\textbf{Model}} & \multicolumn{2}{c}{\textbf{\emph{test-zh}}} & \multicolumn{2}{c}{\textbf{\emph{test-en}}} & \multicolumn{2}{c}{\textbf{\emph{test-hard}}} \\
			\cmidrule(r){2-3} \cmidrule(r){4-5} \cmidrule(r){6-7}
			& \textbf{CER (\%)~$\downarrow$} & \multicolumn{1}{c}{\textbf{SS~$\uparrow$}} & \textbf{WER (\%)~$\downarrow$} & \multicolumn{1}{c}{\textbf{SS~$\uparrow$}} & \textbf{CER (\%)~$\downarrow$} & \multicolumn{1}{c}{\textbf{SS~$\uparrow$}}  \\
			\midrule
                \multicolumn{7}{c}{\textbf{3000-hour Dataset}} \\
                \midrule
                \textbf{SoundStream($1^{st}$ VQ)}~\cite{DBLP:journals/taslp/ZeghidourLOST22} & 14.19 & 0.457 & 25.34 & 0.301  & 27.05 & 0.455 \\
                \textbf{HuBERT}~\cite{hsu2021hubert} & 18.68 & 0.716 & 6.50 & 0.609  & 33.83 & 0.699 \\
                \textbf{W2v-BERT 2.0}~\cite{chung2021w2v} & 2.62 & 0.381 & 6.72 & 0.261  & 23.89 & 0.374 \\
                \textbf{CosyVoice 2.0}~\cite{du2024cosyvoice2} & 1.92 & 0.668 & 7.21 & 0.535 & 15.99  & 0.645 \\
                \textbf{CosyVoice 3.0-0.5B} & 1.68 & 0.710 & 6.60 & 0.614  & 27.60 & 0.679 \\
                
                \midrule
                \multicolumn{7}{c}{\textbf{170,000-hour Dataset}} \\
                \midrule
                
                \textbf{CosyVoice 2.0}~\cite{du2024cosyvoice2} & 1.45 & 0.806 & 2.57 & 0.736  & \textbf{6.83} & 0.776 \\
                \textbf{CosyVoice 3.0-0.5B} & \textbf{1.27} & \textbf{0.815} & \textbf{2.46} & \textbf{0.747}  & 6.96 & \textbf{0.787} \\
			\bottomrule
	\end{tabular}}
	\caption{Performance comparison of down-streaming zero-shot TTS modeling using different tokenizers on the SEED test sets in terms of content consistency (WER/CER) and speaker similarity (SS). While the \textbf{boldface} denotes the best result.}
	\label{tab:ds-token-test}
\end{table*}

Beyond upstream recognition tasks, we also evaluate the tokenizer in downstream TTS tasks to directly assess synthesis performance by replacing the CosyVoice 3 tokens with others and maintaining the model architectures of LM and CFM unchanged. Table \ref{tab:ds-token-test} presents the results of various models trained on datasets of two different scales: 3,000 hours and 170,000 hours. Alongside our supervised semantic tokenizers, CosyVoice 2.0\footnote{https://github.com/FunAudioLLM/CosyVoice} and CosyVoice 3.0, we evaluate the self-supervised tokenizers, HuBERT\footnote{https://github.com/facebookresearch/fairseq/tree/main/examples/hubert} and W2v-BERT 2.0\footnote{https://huggingface.co/amphion/MaskGCT/tree/main/semantic\_codec} which are widely used in other TTS models. In addition, we also involve the unsupervised tokenizer, SoundStream\footnote{https://huggingface.co/amphion/MaskGCT/tree/main/acoustic\_codec}, which quantizes the acoustic waveform into groups of discrete tokens by the residual vector quantization based variational autoencoder (RVQ-VAE). Since other tokens have only single codebook, only the first VQ group of SoundStream is employed for comparison. 

On the 3,000-hour dataset, supervised semantic tokenizers exhibit similar speaker similarity to HuBERT while significantly outperforming W2v-BERT 2.0. This is because both HuBERT and supervised semantic tokenizers focus on semantic information, minimizing acoustic interference, whereas W2v-BERT 2.0 retains all contextual information, both semantic and acoustic, due to its training approach. This allows the conditional flow matching model to better emphasize the acoustic characteristics of reference speech while disregarding acoustic interference in speech tokens. Regarding content consistency, supervised tokenizers achieve the lowest CER on the test-zh set and deliver comparable performance on the test-en and test-hard sets. The notably high CER of HuBERT on the test-zh set underscores its language-specific limitations. As expected, acoustic tokens of SoundStream achieve notably high error rates on all evaluated test sets, indicating poor content consistency to the synthesis text. This is because these acoustic tokens neither attempt to model contextual information like self-supervised tokens nor align with the text like supervised tokens, resulting in a lack of sufficient semantic information.

Increasing the training data volume from 3,000 to 170,000 hours leads to significant improvements in content consistency and speaker similarity, especially for English and challenging scenarios, with relative WER/CER improvements ranging from 63\% to 75\%. As indicated in Table \ref{tab:tts-seed-test}, further scaling the dataset to one million hours enhances performance, but the rate of improvement begins to plateau. This suggests that our multi-task supervised tokenizer is scalable and benefits from larger datasets up to a point of diminishing returns.

\subsection{Ablation of Reinforcement Learning}

Our experiments demonstrate that DiffRO significantly enhances the performance of TTS systems, including both CosyVoice 2 and CosyVoice 3. As indicated in Tables \ref{tab:tts-seed-test}, \ref{tab:ml-clone}, and \ref{tab:cl-clone}, DiffRO achieves relative improvements ranging from 20\% to 50\% in terms of WER. The enhancements are particularly notable in low-resource languages and cross-lingual scenarios, with over 50\% relative WER improvement in half of the conditions; notably, CosyVoice 3-0.5B shows a 68.7\% relative improvement in Korean. Furthermore, RL training reduces the performance gap between the 0.5B and 1.5B models.

Regarding speaker similarity, RL slightly reduces speaker similarity across most datasets, although the change is minimal. This suggests the persistence of the "hacking" problem in DiffRO, where the model focuses more on target rewards, potentially neglecting other metrics. Introducing a speaker similarity module as a reward task may mitigate this issue but could increase WER. Additionally, incorporating the SER task as a reward model aims to enhance CosyVoice 3's emotion expression. Table~\ref{tab:emo} shows that DiffRO-EMO allows CosyVoice3-1.5B to achieve top emotion accuracy across most emotions in both text-related and text-unrelated tasks. However, this improvement in emotion expression can adversely affect pronunciation, highlighting the challenge of balancing rewards in DiffRO, which will be addressed in future work.

Moreover, as seen in Table~\ref{tab:hard-ml-clone}, DiffRO's improvements in WER, SS, and DNSMOS on hard sample test sets are less pronounced than those on overall test sets. This is likely due to the hard samples comprising rare words, tongue twisters, and repeated words, which present significant challenges for reward models.

\subsection{Pronunciation Inpainting}
We construct an evaluation set to compare different pronunciation inpainting methods, focusing on challenging cases of Chinese polyphonic characters and English polyphonic words. Correction rate serves as the metric for assessing inpainting capability. As shown in Table 10, the best method achieves a 100\% correction rate.

The ``RepAll'' approach involves considering all Chinese characters and English words as potential replacements, using internal G2P models for phoneme prediction during training data augmentation. While this method offers extensive coverage of character-phoneme combinations, it introduces mismatches due to G2P predictions. Conversely, ``RepMono'' only replaces monophonic characters or words, ensuring accuracy in the training set.

The key distinction between ``CatPhn'' and ``MixPhn'' lies in whether the Chinese character is retained and concatenated with its phoneme representation or replaced solely by the phoneme. ``CatPhn'' preserves semantic completeness but requires the model to prioritize phoneme representation over the character, which is exacerbated when only monophonic characters are considered. To mitigate this, we introduce some noisy data, such as replacing a character with a different-sounding one while retaining the correct phoneme representation. However, achieving a competitive correction rate with ``MixPhn'' remains challenging.

\begin{table}[t]
\centering{%
\setlength\tabcolsep{4.5pt}
\begin{tabular}{l|ccc|ccc}
\toprule
\multirow{2}{*}{\textbf{Method}}  & \multicolumn{3}{c|}{\textbf{zh}} & \multicolumn{3}{c}{\textbf{en}} \\ 
  & \textbf{Errors} & \textbf{Corrections} & \textbf{Rate(\%)} & \textbf{Errors} & \textbf{Corrections} & \textbf{Rate(\%)} \\ 
\midrule
\textbf{RepAll + MixPhn} & 13 & 9 & 69.2 & 11 & 8 & 72.7 \\
\textbf{RepMono + MixPhn}  & 15 & 15 & 100 & 9 & 9 & 100 \\
\textbf{RepMono + CatPhn} & 15 & 13 & 86.7 & 8 & 8 & 100 \\

\bottomrule
\end{tabular}
}
\vspace{0.05cm}
\caption{Corretion Rate of procunciation inpainting.}
\label{tab:cr_pi}
\end{table}

\subsection{Instructed Generation}
We evaluate the effectiveness of instructed generation capabilities using the Expresso~\cite{expresso} dataset alongside an internal expressive dataset. The Expresso dataset is a multi-speaker expressive speech collection featuring eight distinct speaking styles, evaluated on a subset of 3,000 samples. Our internal dataset includes 3,600 samples, matching the domains of the instruction-following training dataset and encompassing over 50 different emotions, speeds, dialects, accents, and role-playing speaking styles.

The evaluation results are presented in Table~\ref{tab:instructed-test}. CosyVoice 3 shows notable improvements in style similarity, with an approximate 11\% relative increase over its predecessor. In terms of content consistency, CosyVoice 3 demonstrates a higher WER on the Expresso test set but a lower WER on our internal test set. This discrepancy is largely due to the ASR model's bias towards standard pronunciations over emotional ones, as indicated by the higher WER for ground-truth utterances compared to CosyVoice 2. Objectively evaluating content consistency in emotional speech remains a challenging issue.

While we have explored various styles through instructed generation, singing has not been included and will be addressed in future work. Currently, CosyVoice 3's instructed generation focuses on emotion, speech, and style, primarily related to the language model (LM). Timbre, more closely associated with conditional flow matching (CFM), has not yet been considered. Editing timbre using natural language or other modalities is a promising and underexplored area \cite{sheng2025unispeaker}.

\begin{table}[tbh]
\centering{%
\setlength\tabcolsep{12.5pt}
\begin{tabular}{l|ccc|ccc}
\toprule
\multirow{2}{*}{\textbf{Model}}  & \multicolumn{3}{c|}{\textbf{Expresso}} & \multicolumn{3}{c}{\textbf{Internal Dataset}}  \\ 
  & \textbf{WER} & \textbf{SIM} & \textbf{MOS} & \textbf{WER} & \textbf{SIM} & \textbf{MOS} \\ 
\midrule
\textbf{GroundTruth} & 10.0 & 100 & 3.65 & 8.98 & 100 & 3.47 \\
\textbf{CosyVoice 2} & 9.42 & 60.98 & 3.54 & 7.75 & 72.99 & 3.53 \\
\textbf{CosyVoice 3-0.5B}  &  13.72 & 67.82 & 3.56 & 7.30 & 80.45 & 3.51 \\
\textbf{CosyVoice 3-1.5B} &  13.43 & 68.25 & 3.56 & 7.31 & 81.06 & 3.51 \\

\bottomrule
\end{tabular}
}
\vspace{0.05cm}
\caption{Comparison of WER (\%), Style Similarity (SIM), and MOS scores across different models for instructed TTS tasks.}
\label{tab:instructed-test}
\end{table}

\subsection{Results on Speaker Fine-tuned Models}
\begin{figure}[htbp]
    \centering
    \includegraphics[width=1.0\textwidth]{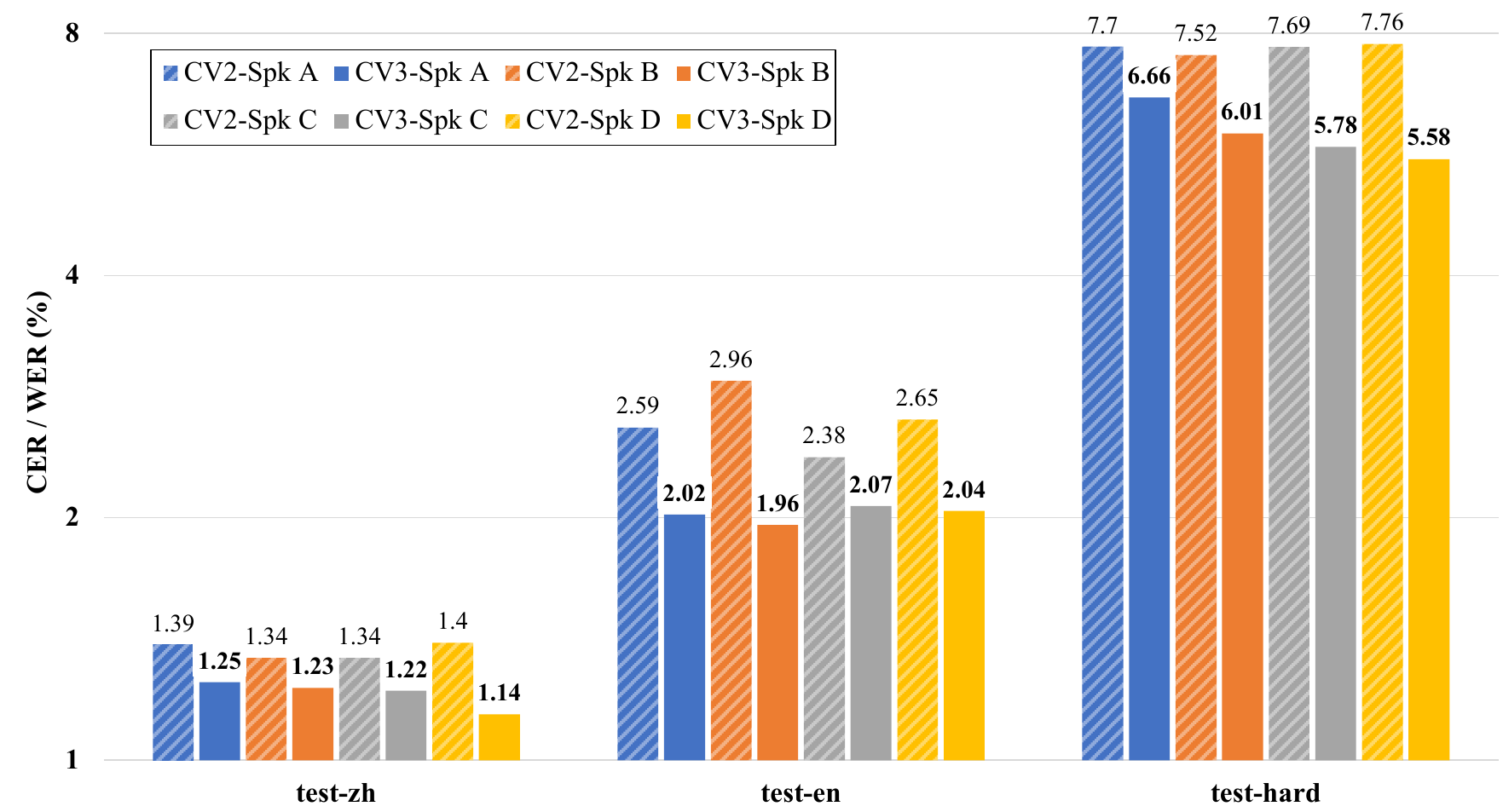}
    \caption{Content consistency results of CosyVoice 3 and CosyVoice 2 SFT models under the SEED-TTS-Eval settings. Word error rate (WER) is used for test-en set, while character error rate (CER) is used for the others.}
    \label{fig:SFT}
\end{figure}

To ensure timbre consistency in the SFT models, we utilize an unsupervised clustering method to identify the timbre centers for each speaker. These clustering centers are then used as speaker embeddings in the conditional flow matching model. As illustrated in Figure \ref{fig:SFT}, increasing the volume and diversity of training data, along with upgrading speech tokens, leads to a reduction in error rates for the fine-tuned models, particularly noticeable in the test-en and test-hard sets. This indicates that improving the base model can also benefit the speaker fine-tuned models.

\subsection{Results on Turning a Monolingual Speaker into a Polyglot}
\begin{figure}[tbh]
    \centering
    \includegraphics[width=0.9\textwidth]{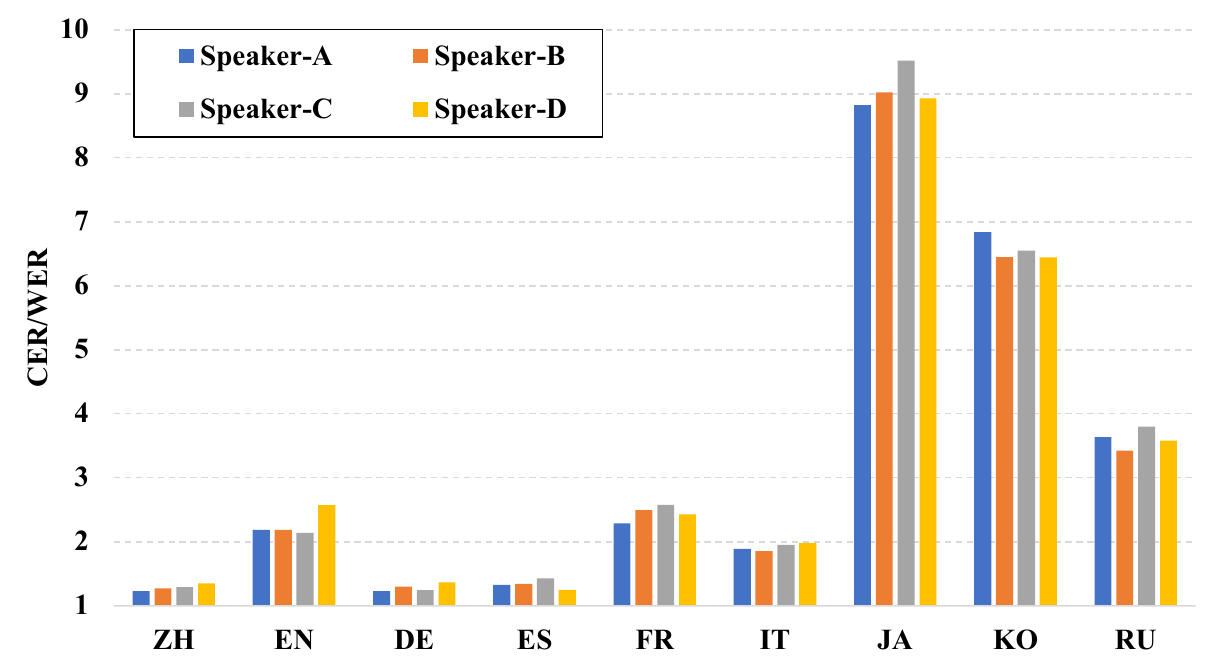}
    \caption{Content consistency results of CosyVoice 3 for turning a monolingual speaker into a polyglot. Character error rate (CER) is used for ZH, KO, and JA, while word error rate (WER) is used for the others.}
    \label{fig:SFT-mlg}
\end{figure}

In our experiments, we aim to transform a monolingual speaker into a polyglot using the training process described in Section \ref{sec:polyglot}. As shown in Figure \ref{fig:SFT-mlg}, the CERs/WERs for languages such as Chinese, English, German, Spanish, French, Italian, and Russian are all below 4\%, demonstrating the effectiveness of our continual training approach.

However, Japanese poses a challenge with a higher character error rate of 9\%, which can be attributed to two main factors: the conversion of kanji into kana before speech synthesis introduces additional errors, and the multiple pronunciations of Japanese characters add complexity. For Korean, the CER is approximately 6\%, mainly due to the limited volume and quality of available data. We will extend the Korean data in the future work.

\section{Conclusion}
To conclude, this report introduces CosyVoice 3, an advanced zero-shot speech synthesis model tailored for in-the-wild applications. By scaling up both data and model parameters, CosyVoice 3 overcomes previous limitations in language coverage and synthesis quality, delivering superior content consistency, speaker similarity, and prosody naturalness. Our innovations, including a novel speech tokenizer and post-training strategies, enhance the model's ability to capture intricate paralinguistic details. Achieving state-of-the-art results across multiple benchmarks, CosyVoice 3 represents a significant step forward in speech synthesis, paving the way for more versatile and high-quality voice generation in diverse real-world scenarios.

\section{Limitations}
CosyVoice 3 has several limitations that need to be addressed in future work. CosyVoice 3 cannot control acoustic characteristics, such as timbre, through textual instructions, which could be an interesting and valuable area of exploration for role-playing applications. Furthermore, CosyVoice 3 does not perform quite well for generating singing voice. This could be improved by adding singing data into the training stages of both tokenizer and LM model.

\bibliographystyle{unsrt}
\bibliography{ref}

\appendix

\end{document}